\documentclass[useAMS,usenatbib,a4paper]{mn2e}
\usepackage{epsfig,graphicx,latexsym,amsmath,amssymb}
\usepackage{natbib}
\usepackage[draft]{hyperref}
\usepackage{mathrsfs}
\usepackage{lastpage}

\def\aap{A\&A}%
\def\apj{ApJ}%
\def\apjl{ApJ}%
\def\mnras{MNRAS}%

\usepackage{color}

\bibpunct[,]{(}{)}{;}{a}{}{,}

\def\Sgr{ Sgr A* }

\title[Invisible Bright Giants]
      {Making Bright Giants Invisible At The Galactic Centre}

\author[Pau Amaro-Seoane et al.] 
{Pau Amaro-Seoane$^{1}$
                        \thanks{E-mail: pau@ice.cat (PAS)},
Xian Chen$^{2}$, Rainer Sch{\"o}del$^{3}$ \& Jordi Casanellas$^{4}$
   \\
$^{1}$
Institute of Space Sciences (ICE, CSIC) \& Institut d'Estudis Espacials de Catalunya (IEEC)\\
at Campus UAB, Carrer de Can Magrans s/n 08193 Barcelona, Spain\\
Kavli Institute for Astronomy and Astrophysics, Beijing 100871, China\\
Institute of Applied Mathematics, Academy of Mathematics and Systems Science, CAS, Beijing 100190, China\\
Zentrum f{\"u}r Astronomie und Astrophysik, TU Berlin, Hardenbergstra{\ss}e 36, 10623 Berlin, Germany
\\
$^{2}$
Astronomy Department, School of Physics, Peking University, 100871 Beijing, China\\
Kavli Institute for Astronomy and Astrophysics, Peking University, Beijing 100871, China\\
(Corresponding author.)
\\
$^{3}$
Instituto de Astrof\'isica de Analuc\'ia, Glorieta de la Astronom\'ia s/n, 18008 Granada, Spain
\\
$^{4}$
Dataguda, Septimania 41, Barcelona, Spain
}

\begin{document}

\date{draft \today}

\pagerange{\thepage--\pageref{LastPage}} \pubyear{2018}

\maketitle

\label{firstpage}

\begin{abstract}
Current observations of the Galactic Center (GC) seem to display a core-like
distribution of bright stars from $\sim 5\arcsec$ inwards.
On the other hand, 
we observe young, massive stars at the GC, with roughly 20-50\% of them in a disc, mostly in the region where the bright giants appear to be lacking.
In a previous publication we put the
idea forward that the missing stars are deeply connected to the presence of this disc.  The progenitor of the stellar disc is very likely to have been a gaseous disc that
at some point fragmented and triggered star formation. This caused the
appearance of overdensity regions in the disc that had high enough densities to
ensure stripping large giants of their atmospheres and thus rendering them very faint.
In
this paper we use a stellar evolution code to derive the properties that a red
giant would display in a colour-magnitude diagram, as well as a
non-linearity factor required for a correct estimate of the
mass loss.
We find that in a very short timescale, the red giants (RGs) leave their standard
	evolutionary track. The non-linearity factor has values that 
	{not only} depend 
	on the properties of the clumps, but {also 
	on the physical conditions the giant stars}, as we
predicted analytically. 
According to our results, envelope stripping works, moving stars on a
short timescale from the giant branch to the white dwarf stage, thus rendering
them invisible to observations.
\end{abstract}


\section{Introduction}
\label{sec.intro}

Stellar dynamics predicts that the stars inside the influence radius of a
supermassive black hole (SMBH) will relax in the energy and angular-momentum
space to form a cuspy distribution with the stellar density increasing towards
the SMBH
\citep{peebles72,BahcallWolf76,ASEtAl04,freitag06,alexander09,PretoAmaroSeoane10,Amaro-SeoanePreto11,Amaro-SeoaneLRR2012}.

These theoretical findings have been thought to contradict observations of the
GC, which seemed to show that the surface density of bright stars\footnote{In
this context, ``bright'' refers to giant stars as luminous or more luminous as
Red Clump stars, which have an observed extinguished K magnitude at the GC of
about 15-16.} within a distance of $0.5~{\rm pc}$ from the SMBH Sgr A* is
constant (``core-like'') if not decreasing towards the SMBH (``hole-like''
\citealt{genzel96,buchholz09,do09,yusef12}).  However, recently, the works of
\citet{Gallego-CanoEtAl2018} and \citet{Schoedeletal2018} have addressed the
observational data at the GC with an improved methodology which allowed them to
obtain star counts for fainter stars than in previous workds and to obtain the surface brightness profile of the
diffuse stellar light. As a result of their analysis, they find that the
stellar density can be described by a three-dimensional power-law with an
exponent value of $\approx -1.2$. Remarkably, these results are in very good
agreement with the numerical work of \cite{BaumgardtEtAl2018}. However,
they also confirm the lack of bight giants within $\sim 0.2\,\textrm{pc}$
of SgrA*. Also, the work of \cite{HabibiEtAl2019} finds a cusp in the old
giants population within $0.02-0.4\,\textrm{pc}$.

Several dynamical processes have been proposed to explain the difference of the distribution of
red giants (RGs) from other stars.
{It was suggested that collisions with main-sequence stars could
reduce the density of RGs \citep{genzel96}, but 
later calculations showed that the resulting size of the core-like nucleus would be too small to explain the
observations
\citep{davies98,alexander99,bailey99,dale09}.
}
More recently, it was found that the collision of RGs with dense gas clumps in a
star-forming gaseous disc, such as the one that formed the observed stellar
disc in our GC, { could more efficiently destroy RGs and 
reproduce the core size} \citep[see][``Paper I'' from now onwards]{pau14}.
Compared to
main-sequence, RGs have larger sizes and lower
surface gravity, and hence are more susceptible to mass loss during such a
collision \citep[see][ who addressed this from a numerical standpoint in the
context of homogeneous discs]{armitage96}.  Moreover, the amount of mass that
is lost grows exponentially if the collision repeats (``non-linearity''),
because after each collision a RG inflates and the self-gravity of the outer
envelope decreases (as explained in Paper I, but see as well
\citealt{armitage96,kieffer16}).  The condition in the GC is favorable to multiple
collisions, as explained in Paper I, because in the central $0.5~{\rm pc}$ are
$\sim 200$ massive young stars \citep{genzel10}, indicating that at least an
equal amount of gas clouds had existed in the past\footnote{ \cite{kieffer16}
found that multiple collisions are unlikely.  This is due to a very small cloud
size they derived from a thin accretion-disc model. But observations indicate
that the accretion disc is mildly thick, so that the resulting cloud size is
much larger (see Paper I for the derivation).}.

In Paper I we presented the idea of how to make giant stars invisible at the
GC.  Due to the limitations of the analytical calculations, a few questions
remain open: (i) a RG which has lost a significant amount of its envelope
could become brighter, or dimmer, depending on its evolutionary stage
\citep{dray06}. This means that RG-cloud collisions do not necessarily remove
RGs from the sensible range of a telescope, and hence this not necessarily
solves the aforementioned ``missing RG problem''; (ii) Addressing this issue
requires calculation of the evolution tracks of RGs in the Hertzsprung-Russell
diagram \citep[pointed out by][]{kieffer16}; (iii) The estimation of the
non-linearity factor $f$, necessary to calculate the mass loss per
hit, is difficult, if not impossible, to derive from first principles and (iv)
the properties of the released degenerate cores {were not
addressed in Paper I}.

\section{Method}

In this section we present the algorithm to calculate the partial removal of
the envelope of a bright giant after hitting a clump, which is the scenario
presented in Paper I, as well as the numerical stellar code and the
implementation of the mass loss in it.

\subsection{Mass loss during collisions}
\label{subsec.massloss}

When a red giant (RG) collides with a gas clump of surface density $\Sigma_c$
and mass $M_c$ at a relative velocity of $v_c$, only that part of the envelope
satisfying the condition

\begin{equation} \label{eqn:cond}
\Sigma_*(R)\sqrt{\frac{GM_*(R)}{R}}<\Sigma_c v_c
\end{equation}

\noindent
will be stripped off the RG \citep{armitage96}.  Here, $\Sigma_*(R)$ is the surface
density (see below for derivation) of the RG as a function of stellar radius
$R$, and $M_*(R)$ is the mass enclosed inside $R$.  For a collision happening
at a distance $D$ from the \Sgr, $v_c\simeq 400 [D/(0.1~{\rm pc})]^{-1/2}~{\rm
km~s^{-1}}$.  In the following we adapt $v_c=400~{\rm km~s^{-1}}$ as a fiducial value
for our simulations.

Following the analysis presented in Paper I, and using the same notation, we calculate
$\Sigma_c$ with

\begin{align}
\Sigma_c\simeq 2\cdot\,10^{4}~{\rm
g~cm^{-2}}\left(\frac{M_c}{10^2~M_\odot}\right)^{-1}\left(\frac{D}{0.1{\rm
pc}}\right)^{-2}\left(\frac{H}{0.1D}\right)^4.\label{eqn:sigc}
\end{align}

\noindent
Assuming that the accretion disc which created the clumps has a geometric
thickness of $H/D=0.1$ {and the typical mass of the clumps 
is $10^2\,M_\odot$}, we estimate that $\Sigma_c=2\cdot\,10^{4}~{\rm
g~cm^{-2}}$ for a clump at $D=0.1$ pc. 
{For these parameters, the size of the clumps is typically $R_c\sim10^{-2}D$.
Given that today there are about $10^2$
massive stars in the observed young stellar disc in the GC, the total mass
of the disk is about $10^4\,M_\odot$, which is also consistent with the mass of the disc
required by the dynamical model to explain the distribution of the other young stars
(e.g., S-stars) in the GC
\citep{chen14}.}

The mass loss during one collision, $\Delta M$,
is calculated as follows with Equation~(\ref{eqn:cond}):

\begin{enumerate}

\item Derive the function $M_*(R)$ from the output of the stellar evolution
model. If not in the output, calculate it using the density $\rho_*(r)$ of the
RG,

\begin{equation}
M_*(R)=4\pi\int_{0}^{R}\rho(r)r^2dr.
\end{equation}

\item Derive $\Sigma_*(R)$ for any $0<R<R_*$, where $R_*$ is the radius of the
RG,

\begin{equation} \label{eqn:sigmaint}
\Sigma_*(R) =2\int_{0}^{\sqrt{R_*^2-R^2}} \rho(\sqrt{h^2+R^2}) dh.
\end{equation}

\item Find the critical radius $R_s$, such that Equation~(\ref{eqn:cond}) is
satisfied when $R>R_s$.

\item The stellar mass at $R>R_s$ that will be stripped off the RG during the
collision, is therefore

\begin{equation} \label{eqn:masslosscoll}
\Delta M=M_*(R_*)-M_*(R_s).
\end{equation}

\end{enumerate}

\subsection{Implementation in a stellar evolution code}

We implement this mechanism of mass loss in the stellar evolution code CESAM
\citep{2008Ap&SS.316...61M}. The code allows one to compute stellar
models from the pre-main sequence phase to advanced stages.
It has been widely used in the context of asteroseismic studies
\citep{2009ApJ...700.1589S} and in particular to simulate red-giant stars
\citep{2012A&A...540A.143M,2011A&A...526A.100P}.\\

We modify the original subroutines for external, constant mass
loss to include the mass loss due to repeated collisions with a gas clump. For
each collision, we solve Equations ~\ref{eqn:cond}, \ref{eqn:sigmaint} and
\ref{eqn:masslosscoll} to find the critical radius $R_s$ above which all mass
is stripped from the star. The amount of mass lost in each collision, $\Delta
M$, is spread over a time span of $10^3$ yr to guarantee the convergence of the
code in the face of the strong structural variations that occur when the
envelope is stripped. 

We simulate the evolution through the red giant branch of stars from
$M_{\star}=1\;\textmd{M}_{\odot}$ to 7~M$_{\odot}$ (note that that high masses
are of interest to also consider recent star formation; for a population older
than $1\,(3)$ Gyr, it is enough to consider masses $<\,3\,(2)\,M_{\odot}$).  We
then consider collisions with gas clumps with surface densities from $\Sigma_c
= 10^{4}$~g~cm$^{-2}$ to $10^{6}$~g~cm$^{-2}$. The collisions with the gas
clumps start when the stars evolve through the red giant branch, once they
reach radii ranging from $R_{\star, i}=30\;\textmd{R}_{\sun}$ to
100~R$_{\sun}$. After removing the mass lost in one collision, we let the stars
evolve during $5\cdot 10^4$ yr without any mass loss before undergoing another
collision.  {This treatment leads to $20$ collisions in every $1$ Myr,
to be consistent with the estimations in Paper I.
The time during which the stars experience multiple collisions depends on the
timescale of the fragmenting phase of the stellar disc, which in turn depends
on the lifetime of the clumps. In Paper I we found that the typical lifetime is
$10^5$ years but for less massive clumps (i.e., denser clumps according to
Eq.~\ref{eqn:sigc}) the lifetime could be as long as a few Myrs. Therefore, we
vary the collision time between $0.4$ and $5$ Myrs.}

\section{Results}

\begin{figure}
\resizebox{\hsize}{!}
          {\includegraphics[scale=1,clip]{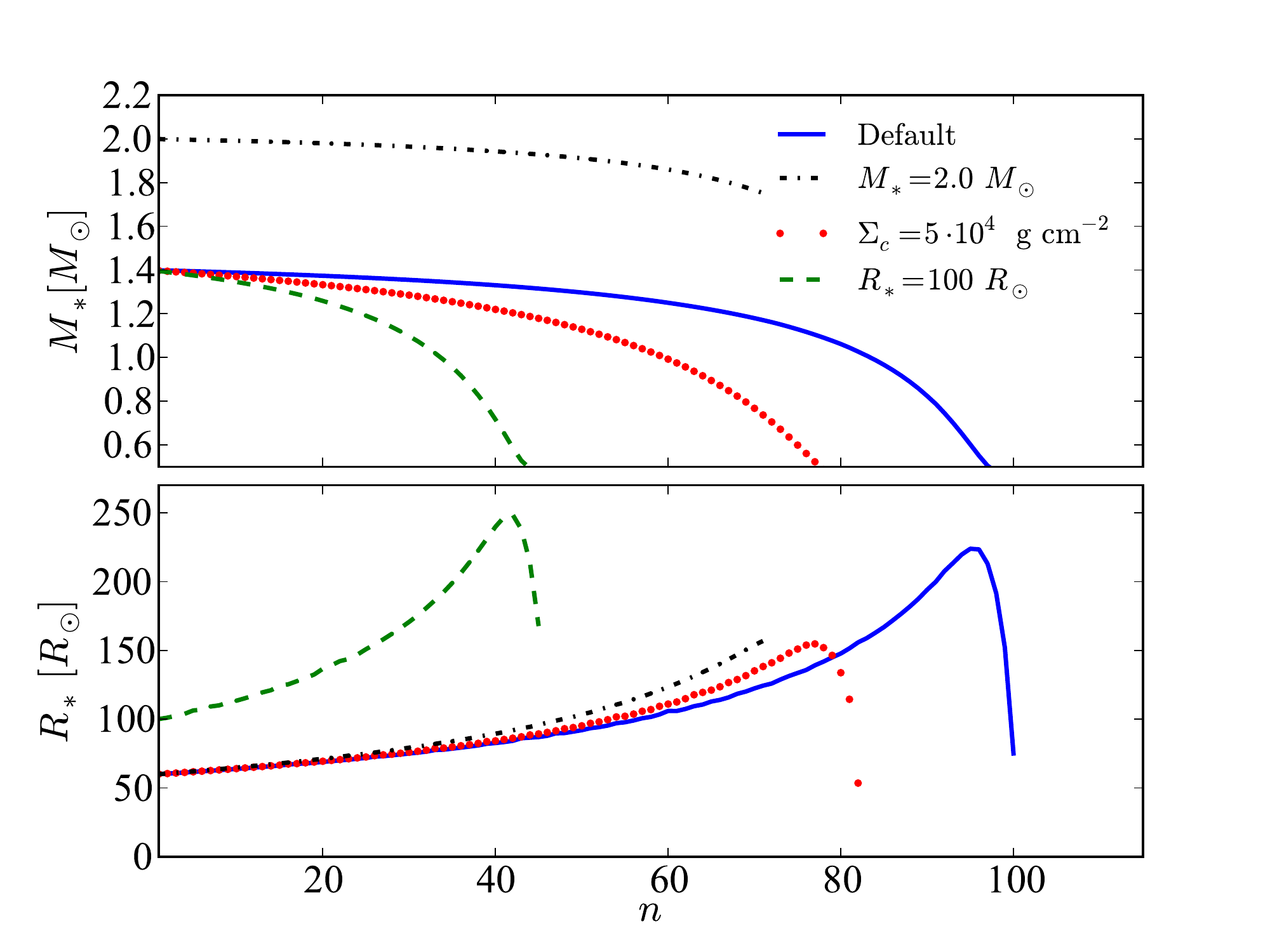}}
\caption
   {
Evolution of the mass $M_*$ (upper panel) and the radius of the star $R_*$
(lower panel) as a function of the number of hits $n$ with the clumps in the
star-forming disc.
The model parameters by default are $M_*=1.4~M_\odot$, $R_*=60~R_\odot$
 and $\Sigma_c=2\times10^{4}~{\rm g~cm^{-2}}$, unless mentioned otherwise in
the legend.
} \label{fig.M_All}
\end{figure}

In Figure~\ref{fig.M_All} we depict the evolution of the stellar masses and
radii for red giants as a function of the number of accumulated hits with
clumps. The larger the radius, the more efficient the removal of mass from the
envelope is. Obviously, the more massive the star is, the longer it takes to
achieve a significant mass loss.  {The figure also shows that during the
first collisions, the loss of the envelop leads to an expansion of the stellar
radius. The expansion is caused by the higher gas pressure in the deeper part
of the envelope. When almost all the envelope is removed,the star stops
to expand and shrinks after each collision.} 

\begin{figure*}
\resizebox{0.9\hsize}{!}
          {\includegraphics[scale=0.9,clip]{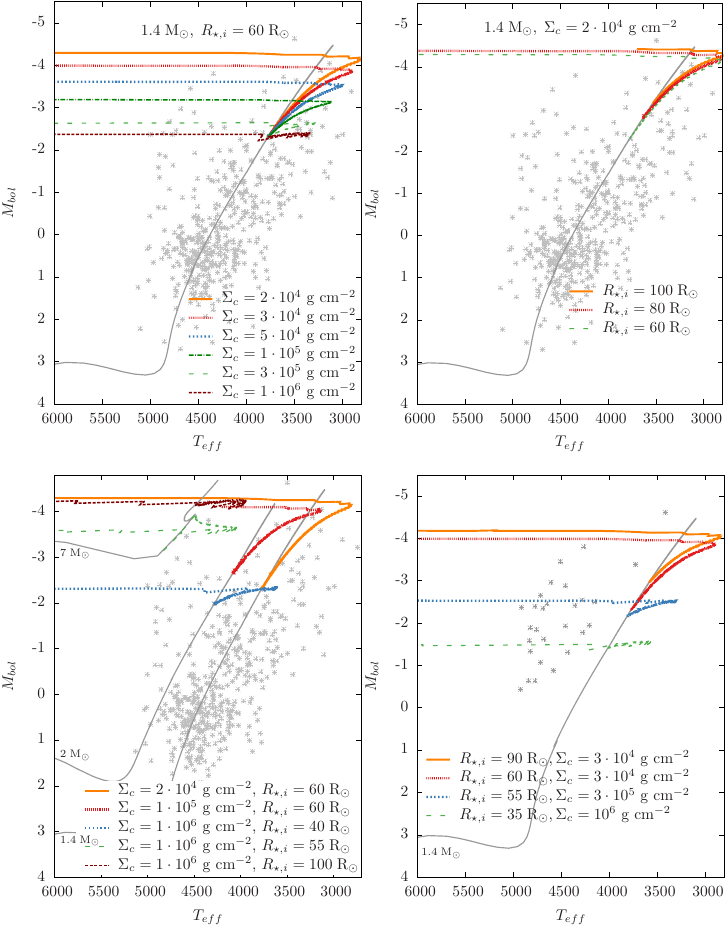}}
\caption
   {
Hertzsprung-Russell (HR) diagram (Bolometric magnitude $M_{bol}$ against
	effective temperature $T_{eff}$) for different bright giants and
properties of the clumps, including (a) a fixed mass and radius of the star
and six different surface densities of the clumps (top left),
(b) a fixed mass and surface density and three different stellar
radii (top right), (c) three different initial masses, four
different initial radii and three different surface densities (lower
left) and (d) one initial stellar mass, four different initial
stellar radii and three different surface densities (lower right).
{The gray solid curves show the standard evolutionary tracks of RGs,
	and the colored curves are derived from our model which inlduces the
	collisions between RGs and gas clumps. The asterisk symbols show the
	distribution of RGs in the GC (Ref. here).}
   }
\label{fig.HR_all}
\end{figure*}

In {Figure~\ref{fig.HR_all} we show the evolution of the RGs in the
Hertzsprung-Russell (HR) diagram.  In general, a RG becomes brighter and cooler
during the first tens of collisions, as a result of the expansion of the
stellar radius.  When almost all the envelope is removed, the evolutionary
track turns horizontally: The star becomes hotter in successive collisions
while the bolometric luminosity remains constant.}

\begin{figure}
\resizebox{\hsize}{!}
          {\includegraphics[scale=1,clip]{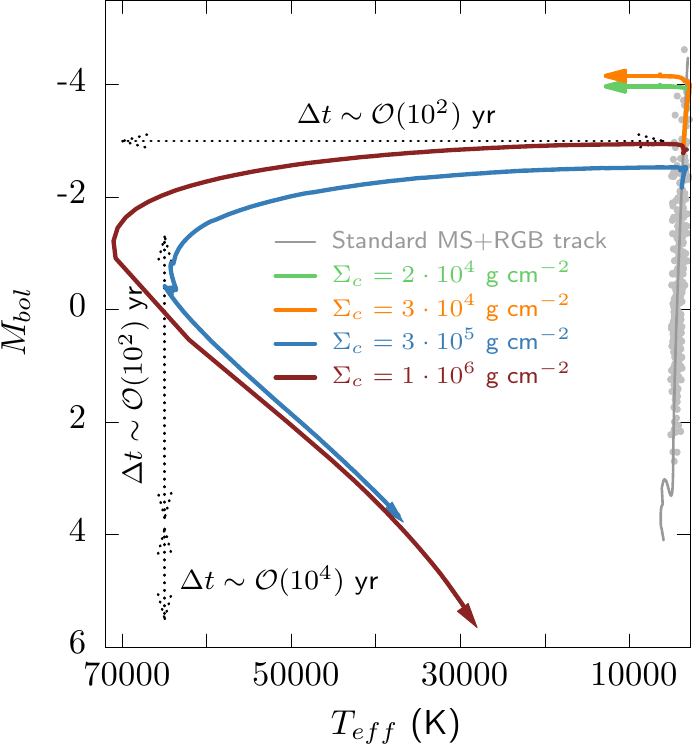}}
\caption
   {
Same HR diagram as in Figure~\ref{fig.HR_all} but for much higher temperatures.
We depict the evolutionary track of a $1.4\,M_{\odot}$ and $60\,R_{\odot}$
star after successive impacts with clumps of different surface densities.
In grey, to the right, we have the same standard MS+RGB track displayed in
the aforementioned Figure. From the ``first Knee Point'' (see text) to the maximum
effective temperatures reached, the evolution happens in very short timescales,
of order $\sim 10^2$ yrs. From the ``second Knee Point'' to the end of the
evolution, going back to lower effective temperatures, the timescale, while
longer than the previous one, is still short, of order $\sim 10^4$ yrs.
   }
\label{fig.HR-Lumfall_irad}
\end{figure}

{After leaving the standard track for RGs, the remnant of the RG quickly
cools and settles down on the white-dwarf branch, as is shown in Figure
\ref{fig.HR-Lumfall_irad}.  White dwarfs are too faint to be detected at the GC with current instrumentation. Thus, a RG will become ``invisible'' after envelope stripping.}

\begin{figure}
\resizebox{\hsize}{!}
          {\includegraphics[scale=1,clip]{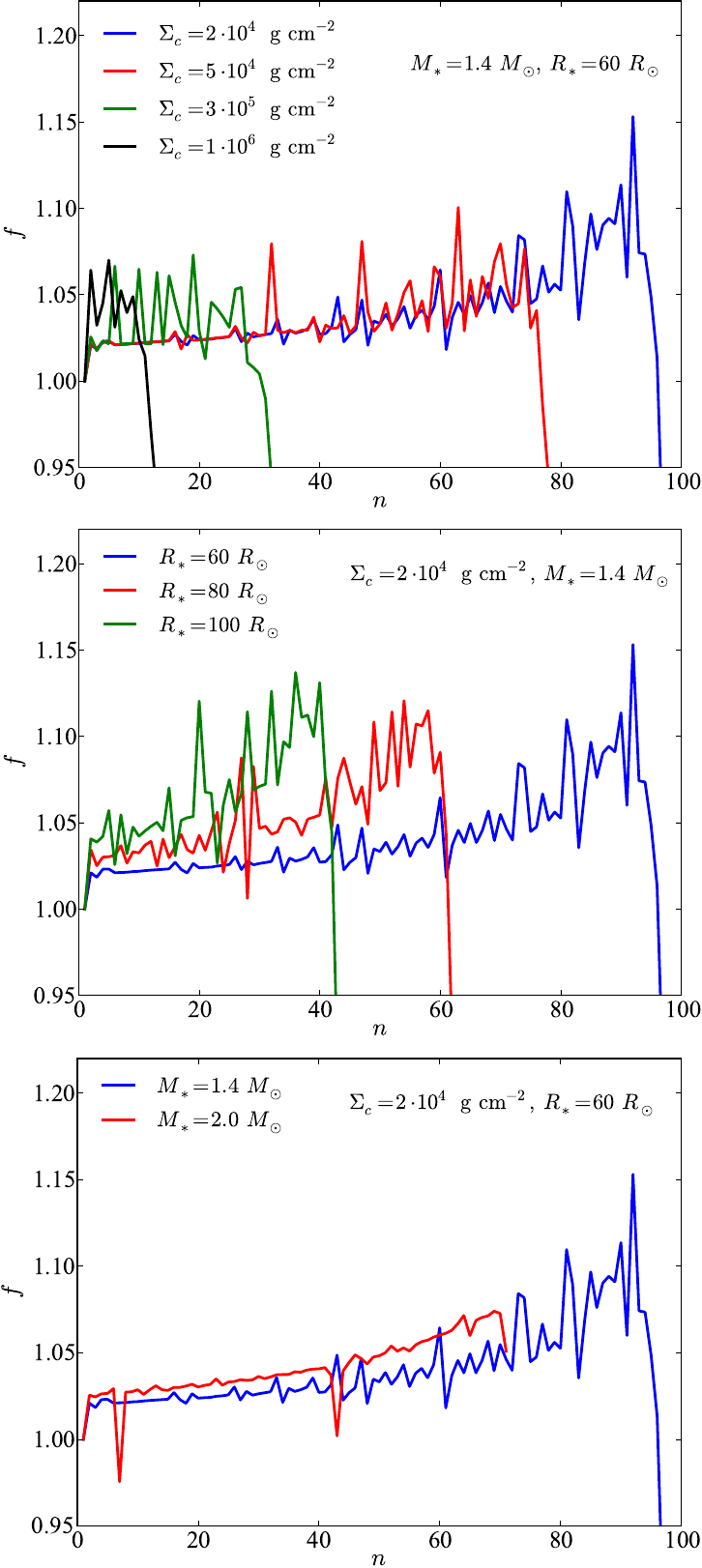}}
\caption
   {
The non-linearity factor, as introduced in Paper I, for four different
surface densities and a fixed stellar mass and radius (upper panel),
three different initial stellar radii, a fixed surface density and
stellar mass (mid panel) and a fixed surface density and stellar radius
and two different initial stellar masses (lower panel).
   }
\label{fig.All_f}
\end{figure}

Using the above method, for each collision we derive the amount of  mass
$\Delta M$ that is lost from the RG envelope. We denote the mass loss during
the $n$th collision as $\Delta M_n$ ($n\ge1$), and for each collision we
calculate the non-linearity factor with $f=\Delta M_{n+1}/\Delta M_n$.
{The bigger the non-linearity factor is, the smaller number of
collisions are needed to remove the envelope of RGs.  Numerical simulations
with large RGs (with a size of $150R_\odot$) suggests that $f$ is $2$
\citep{armitage96}. The value decreases if the size of a RG is smaller.}
Figure~\ref{fig.All_f} summarizes the results from our simulations:  In general
$f-1={\cal O}(0.1)$, as we estimated in Paper I. The value of $f$ increases
with $n$ because of the binding energy of the star decreases.

\section{Conclusions}

Previous observations of giant stars as luminous or more luminous as Red Clump
stars at the Galactic Center appear to indicate a core-like distribution, and
it is important to note that we cannot say whether this also holds for the
fainter stars, such as main sequence stars. More recent observations, though, reconcile with theory and
numerical simulations
\citep{Gallego-CanoEtAl2018,Schoedeletal2018,BaumgardtEtAl2018}.
\cite{HabibiEtAl2019} confirm the results of \cite{Gallego-CanoEtAl2018} and 
\cite{Schoedeletal2018} that there is a cusp of faint giants ($K_s > 15$).
However, this does not change the basic observation that we are missing bright giants.

Theoretically, as explained in the introduction, {the missing
RGs} poses a problem that has
been studied for more than 15 years. In Paper I, \cite{pau14}, {we 
proposed that the only scenario that could lead to an efficient
depletion of the RGs in the GC is the collision of the giant stars with
the clumps that later become the observed stellar disc.  
Although our analytical approach provided a solution to the problem,
a few open questions remained, including the features of the giant stars in the 
colour-magnitude
diagram as they suffer consecutive hits with the high-density clumps, and the
non-linearity factor that determines the lifetime of the RGs.}

In this paper we address the evolution of a giant star in the GC with a
numerical stellar evolution code which we have modified to take into account
the mass loss due to repeated hits against a gas clump. We find that clumps
remove the envelopes very efficiently. In a very short timescale, the RGs leave
their standard evolutionary track and in some $10^5$ years  move to effective
temperatures as high as 70,000K and drop their bolometric magnitude by many
orders of magnitude. While this works out very well for RGs, horizontal branch
(HB) stars have an envelope 100 times denser in surface density, and
consequently require about 100 more impacts to be depleted.  As explained in
Paper I, this means that most of the HB stars will suffer a partial depletion
and only a fraction (those with low inclinations) will have received a
significant envelope damage.

We deem it important to stress that (i) we rely on a single episode of disc
formation, but it is likely that many episodes have happened, which would help
to completely destroy all envelopes, (ii) we assume a population of only 100
clumps, because that is the number of stars that we currently observe in the
stellar disc at the GC, although this is a very conservative lower limit.
Having more clumps obviously helps in the removal of the envelopes as well and
(iii) we consider a single mass ($10^2~M_\odot$) for the clumps, while a broad
spectrum is more realistic. Having less massive clumps helps in the process of
removal of RG envelope, because the surface density of a clump is inversely
proportional to its mass (Paper I).

Our theory predicts that giant stars are ``hiding'': While we cannot see them,
their degenerate cores are populating the GC, building the old, segregated cusp
that a number of independent studies have predicted theoretically.  

\section*{Acknowledgments}

PAS acknowledges support from the Ram{\'o}n y Cajal Programme of the Ministry
of Economy, Industry and Competitiveness of Spain, as well as the COST Action
GWverse CA16104. This work was supported by the National Key R\&D Program of
China (2016YFA0400702) and the National Science Foundation of China (11721303
and 11873022). He is indebted with Marta Masini and Luisa Seoane for their
encouragement and countless hours of remote support during his visit to the
KIAA in Beijing.  The research leading to these results has received funding
from the European Research Council under the European Union's Seventh Framework
Programme (FP7/2007-2013) / ERC grant agreement nr 614922. RS acknowledges
financial support from the State Agency for Research of the Spanish MCIU
through the ``Center of Excellence Severo Ochoa'' award for the Instituto de
Astrof{\'i}sica de Andalucia (SEV-2017-0709).  RS  acknowledges financial
support from the national grant PGC2018-095049-B-C21 (MCIU/AEI/FEDER, UE).

\end{document}